\documentclass[journal=nalefd,manuscript=article]{achemso}
\usepackage{graphicx}    % For graph
\usepackage[T1]{fontenc} % Modern font encoding
\usepackage{float}       % For creating charts, graphs and schemes
\usepackage{helvet}      % Helvetica font for sans serif
\usepackage{mathptmx}    % Times font ("Word-like")

\title{Structural stability and uniformity of magnetic Pt$_{13}$ nanoparticles in NaY zeolite}

\author{Luca Pavan}
\affiliation{Department of Physics, King's College London, London, WC2R 2LS UK}
\author{Cono Di Paola}
\affiliation{Department of Dept of Earth Sciences, University College London, London, WC1E 6BT UK}
\author{Francesca Baletto}
\email{francesca.baletto@kcl.ac.uk}
\affiliation{Department of Physics, King's College London, London, WC2R 2LS UK}

\begin{document}

\begin{abstract}
Based on first-principles simulations, the structural stability and magnetic uniformity of Pt$_{13}$ nanoparticles encapsulated in a NaY zeolite were investigated. Among 50 stable isomers in the gas phase, only 15 could be accommodated into a zeolite pore and severe structural rearrangements occured depending on whether the solid angle at the Pt vertex bound to the supercage was larger than 2 sr (i.e. icosahedron). When van der Waals forces were included, the global minimum was found to be a new L-shaped cubic wire that is unstable in the gas phase. The total magnetization of the encapsulated Pt$_{13}$ decreases due to the stabilization of less coordinated isomers, with the majority of clusters charaterized by a total magnetization of 2 $\mu_B$, while the majority of free clusters exhibit a threefold value.

\end{abstract}

\maketitle

%\section{Introduction}
Within the last few decades, the number of studies related to metallic nanoparticles (NPs) has dramatically increased because of their applicability in catalysis, magnetic and optical devices and nanomedicine \cite{Chen10,Barnard10}. One of the main challenges in nanoscience is addressing the structure-property relationship of a structurally uniform sample of metallic NPs in a well defined size range \cite{Sanchez10,Medel2011}. Among transition metals, encapsulated Pt-nanostructures are playing a central role in the fabrication of electrochemical devices \cite{Guo08}, including lithium batteries \cite{Chen09}, because they drastically enhance their catalytic performance for various chemical reactions of industrial interest \cite{Zecevic13,Koenigsmann11}. Furthermore, Pt-nanoystems, as clusters, nanowires, and thin films, exhibit a magnetic behavior \cite{Sakamoto11,Parsina12,Cono13,Smogunov08}, as a consequence of their broken translation symmetry and reduced coordination number has been observed \cite{Maitra11}. Recent SQUID measurements of Pt-NPs consisting of 13$\pm$2 atoms and encaged in faujasite-Na zeolites show that they have a high spin state between $S=2$ and $S=3$, equivalent to a total magnetic moment of 4-6 $\mu_B$ \cite{Liu06,Liu09},  whereas XMCD studies have indicated a peak total magnetization of 5.9 $\mu_B$ \cite{Bartolome09,Batlle11}. In any event, the sample is not uniformly magnetic; only a small fraction of the nominal Pt-load, specifically 15-20\% as measured by XMCD spectroscopy  and 30\% as measured by SQUID magnetometry are magnetic. 3D electron tomography has provided insights into the local structure and shape of Pt-clusters encapsulated in zeolite materials \cite{Roduner14,Akdogan08}, but the same authors have suggested that a local analysis, as opposed to average characterization, is needed to elucidate the structure-performance relationship. On this regard, atomistic simulation is an extremely powerful tool. Theoretical studies of metallic NPs in porous frameworks such as zeolites, are scarce and are usually limited to one or a few atoms and to few geometrical shapes \cite{Bunau2014,Yadnum13,Huertas13,Boefka09}. Recent density-functional theory (DFT) calculations have suggested a Pt-icosahedron (Ih) as the shape grown inside zeolite pores \cite{Liu06jpcb}. Nevertheless, in the gas phase, this geometry is energetically unfavorable with respect to  the buckled pyramid (BP) \cite{Hu09}, the triangular prism (TP) \cite{Wang07}, and the low-symmetry shape (LOW) proposed by Piotrowski et al. \cite{Piotrowski10}, which is the global minimum. Thus, a complete structural description of Pt$_{13}$-NPs inserted into nanoporous materials is missing but highly desirable because it will provide insights into how the embedding affects the structural stability and then the magnetic nature of the sample. 

In this Letter, the physical properties of a variety of encaged Pt$_{13}$ isomers are calculated by DFT simulations, including Van der Waals dispersion forces (DFT-D). As paradigmatic example of porous material, a 1:4 Al/Si faujasite supercage, H$_{30}$O$_{42}$Al$_{6}$Si$_{24}$, containing two Al atoms per oxygen six-ring sub-structure is used as a matrix.
For the first time, we demonstrate that Pt-NPs@NaY are anchored to the supercage through only one atom and that their structural stability depends strongly on the solid angle at that anchor. When the solid angle is greater than 2 sr, i.e. the Ih, the isomer undergoes severe structural changes. At DFT-D, a new concave structure delimited by several square facets is the optimal geometry, although it is unstable in the gas phase. Among the ensemble of more than 15 isomers that can be inserted into a zeolite pore, the number of highly coordinated and highly magnetic (S>3) clusters decreases from the 70\% in the gas phase to 30\% after the embedding, due to the stabilization of low coordinated shapes. 

\section{Model and Results}
Our numerical approach includes four different steps: accurate sampling of the potential energy surface of Pt-NPs in the gas phase with a focus on shapes with an high coordinate number (CN $>$ 5.3) throughout an iterative metadynamics (iMT); refinement of their structural and magnetic properties within a DFT framework; selection and embedding within a NaY-zeolite pore; and the characterization of the structural and magnetic properties, including van der Waals forces (DFT-D). 

Metadynamics (MT) is an algorithm used to accelerate rare events, where the ionic dynamics is biased by a history-dependent potential built on a set of reaction coordinates called collective variables (CVs) \cite{Laio02}. Commonly used to explore free-energy surfaces, MT has been recently proposed for sampling the energy landscape of small clusters \cite{Tribello11}. 
To ensure a detailed sampling of shapes in the gas phase, an iterative procedure (iMT) employing the coordination mumber (CN) as unique CV is used as in Ref. \cite{Pavan13}. The first starting configurations are well-known geometries, such as LOW, BP, truncated bi-pyramid (TBP), TP, icosahedron (Ih), cubooctahedron (Co) and decahedron (Dh). When a morphology belonging to a new structural basin is obtained, this is quenched to its nearest local minimum and then re-used as starting configuration for a new simulation, in an iterative manner. 
To speed up the structural search, a tight-binding second moment interatomic potetial, able to reproduce the geometrical details of Pt clusters, is applied. Anyway, the relative energy stability of each isomer is calculated according to the Broyden-Fletcher-Goldfarb-Shanno (BFGS) algorithm after ionic relaxation, within a scalar-relativistic spin-polarized DFT framework, as available in the Quantum Espresso package \cite{Giannozzi09}. Scalar-relativistic spin-polarized calculations with a non-linear core corrections and a small as 0.03eV Marzari-Vanderbilt smearing are able to reproduce correctly the geometrical features and the magnetic moment of Pt$_{13}$ with respect to relativistic approaches, including spin-orbit coupling \cite{Blonski11,Piotrowski10}.
In the case of free NPs, the simulation box contains at least 13~\AA\, of vacuum to avoid any spurious interaction between periodic images. While for encaged NPs in the H$_{30}$O$_{42}$Al$_{6}$Si$_{24}$ cage, a cubic cell of approximately 28~\AA\ is used and a full BFGS ionic relaxation is applied to Pt atoms and the three inward O atoms of the six-ring substructure. Both standard DFT and corrected DFT for van der Waals forces (DFT-D), where a semi-empirical damp-dispersion term is introduced \cite{Barone09}, are performed. The NaY supercage does not exhibit any magnetic behavior. More details are reported in the Supporting Material pages S1-S3.

\begin{center}
\begin{figure}[h!]
\hspace{-0.4cm}\includegraphics[scale=0.7]{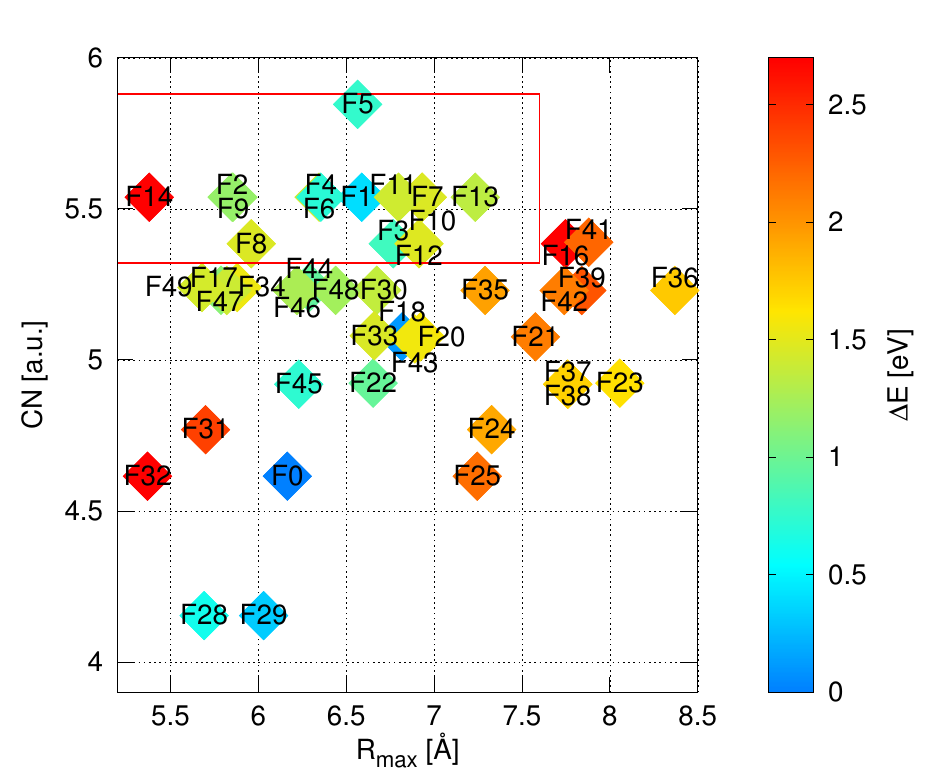}
\caption{Energy stability of Pt$_{13}$ isomers with respect to the global minimum (F0) plotted as a function of their coordination number, CN and maximum pair distance R$_{max}$ in \AA .  The red box features isomers satisfying geometrical requirements.}
\label{Fig_potential}
\end{figure}
\end{center}

%\section{Results}
We first discuss the results for the gas phase. An exhaustive sample of 49 different local minima was obtained by the EP-iMT search followed by the DFT ionic relaxation. In  \ref{Fig_potential}, these isomers are plotted as a function of their average coordination number (CN) and maximum pair distance (R$_{max}$). They are labeled as "F", which represents "free objects", followed by a progressive number from 0 to 48. The color scheme refers to their relative stability, $\Delta E$, with respect to the total energy of the global minimum (F0), which has been confirmed to be the low symmetry Piotrowski's shape \cite{Piotrowski10}.

\begin{center}
\begin{figure*}[h!]
\includegraphics[scale=0.28]{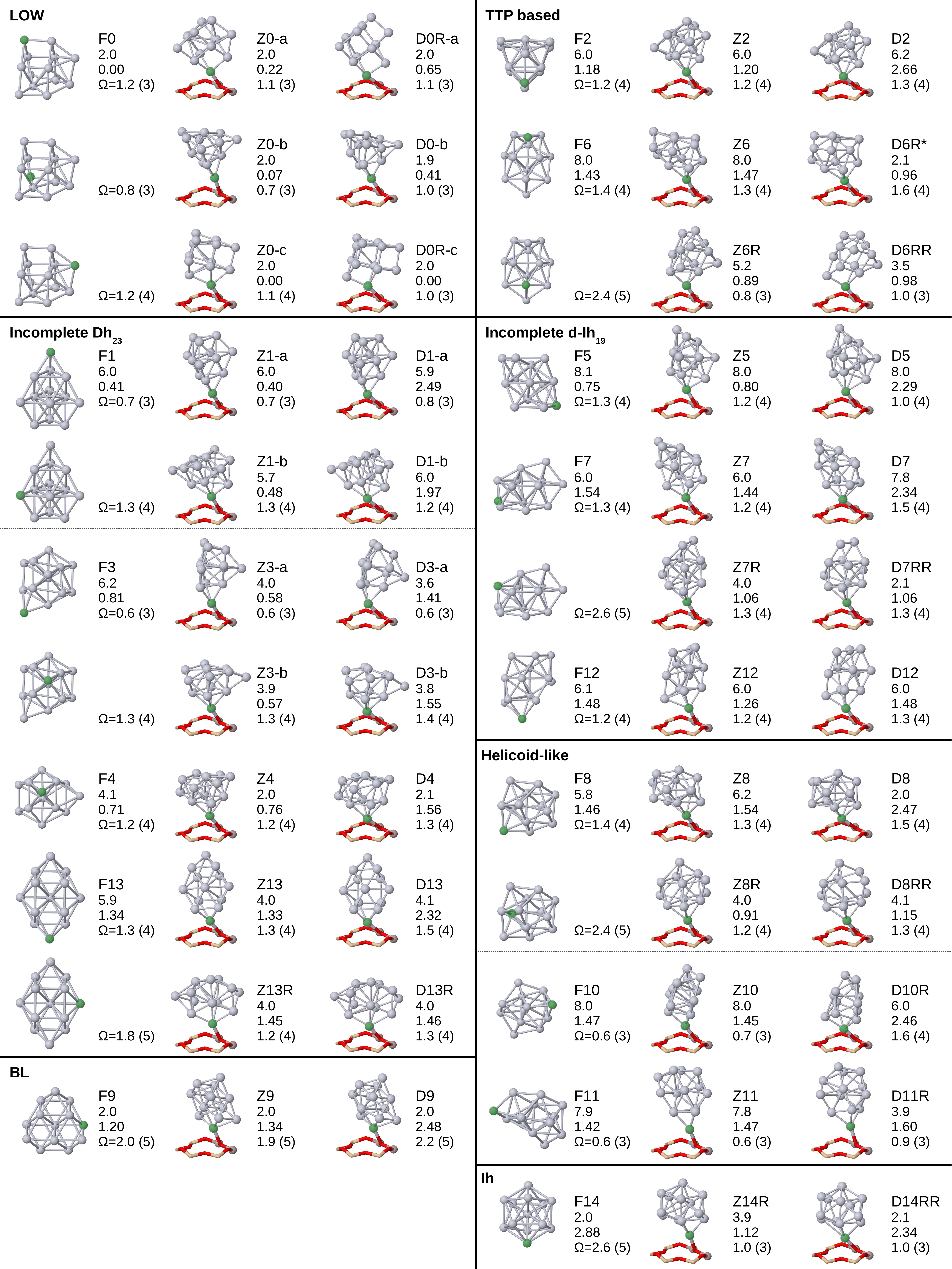}
\caption{Selected Pt$_{13}$ isomers, classified into their geometrical families. The left columns represents the gas phase (F prefixes), the middle column represents Pt$_{13}$ inserted into the zeolite supercage (Z prefixes); and the third column represents refinement at the DFT-D level (D prefixes). For each cluster, the total magnetization in Bohr magneton ($\mu_B$), the relative energy with respect the global minimum (in eV), and the solid angle at the anchor are listed below their identification label. The coordination of the anchor itself is listed in brackets.} 
\label{Fig_selected}
\end{figure*}
\end{center}

\begin{center}
\begin{figure*}
\hspace{-0.6cm}\includegraphics[scale=0.7]{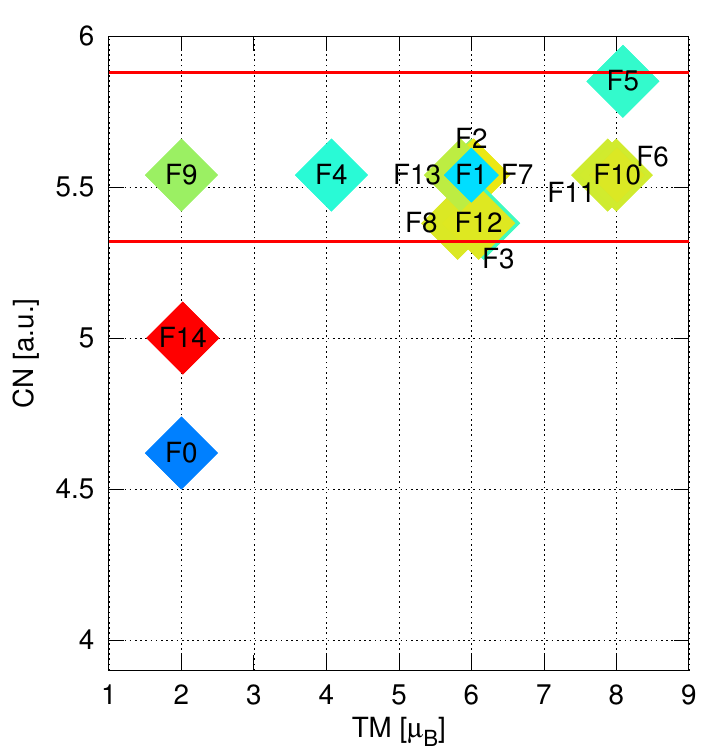} 
\includegraphics[scale=0.7]{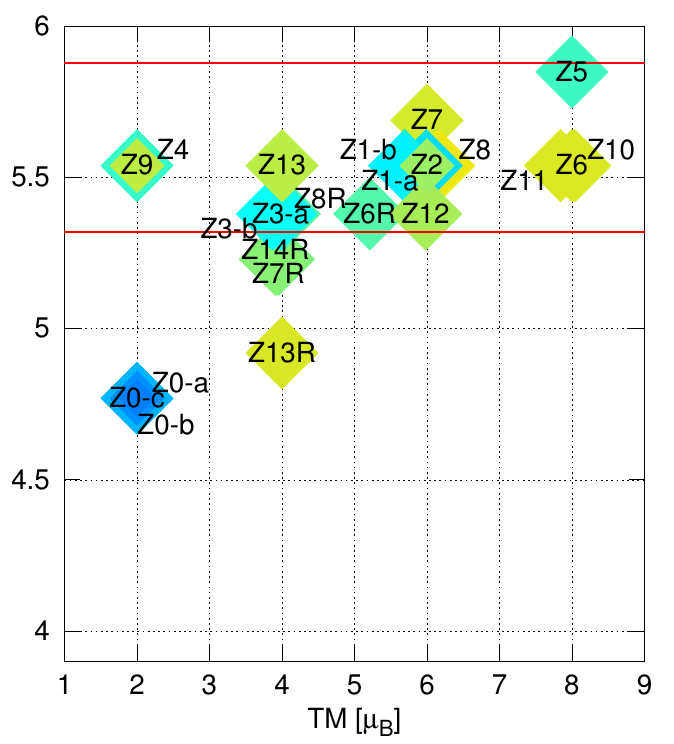}
\includegraphics[scale=0.7]{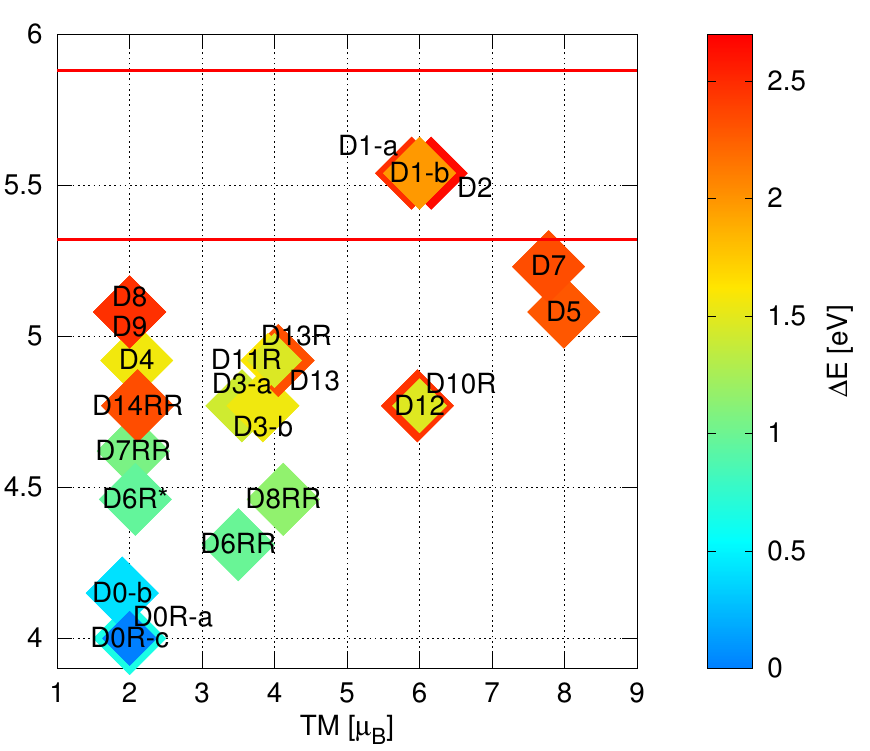}
\caption{Average coordination number (CN) with respect to the total magnetization, TM ($\mu_B$). The color refers to energy stability of
Pt13 -NPs in the gas phase (left), embedded in an NaY zeolite (middle) and in zeolite with dispersion forces DFT-D (right). $\Delta E$ (eV), relative energy stability, to the global minima F0, Z0-c and D0R-c, respectively. The region between the red lines identifies the experimental range of coordination.}
\label{Fig_zeo}
\end{figure*}
\end{center}

The geometrical features of Pt-NPs embedded in NaY-zeolite, as reported in Refs. \cite{Akdogan08,Liu06,Bartolome09}, can be summarized as follows:(i) The coordination number (CN) reaches a peak of 5.6, accordingly to analysis based on  Fourier transform of the EXAFS $\chi$(k) signal; (ii) the chemisorption site is on top of the oxygen six-ring site (SII); and (iii) the diameter should fit the dimensions of the cavity, and therefore be between 6 and 8 \AA, and have a rather spherical shape. The red box in  \ref{Fig_potential} identifies the 14 isomers, F1 to F14, that satisfy those geometrical constraints. A maximum pair distance of 7.5 \AA~ represents the longest distance that can be accommodated in the supercage pore, taking into account the Pt-O bond length and O and Pt van der Waals radii. For the sake of completeness, F0 will be considered in our analysis despite having a very low CN. Notably, the other low-energy isomers, such as the two triangular prisms (F28 and F29), and the buckled pyramid (F18) exhibit CNs that are excessively low with respect to the experimental data and thus will be not considered further.
All the selected clusters are reported in the first column of  \ref{Fig_selected} where are grouped into their geometrical families \cite{Pavan13}. The family of incomplete decahedron of 23 atoms, iDh$_{23}$, has four isomers without fivefold vertex -the TBP (F1), F3, and the stellated half-star shape, F4- and only one with a fivefold vertex (F13). A three-fold symmetric triaugmented triangular prism (TTP), tagged F2, and the Dada-TTP (F6), belong to the same family. Three isomers -F5, F7, and F12- are incomplete double icosahedra of 19 atoms, idIh$_{19}$. Three helicoidal shapes -F8, F10, and F11-, a bi-layer geometry (F9) -the only fully crystallographic local minimum-, and the Ih$_{13}$ (F14) complete the considered ensemble of Pt$_{13}$. Notably, only F1 has a $\Delta E$ below 0.5 eV while all the other isomers, but Ih, lying between 0.6 and 1.5 eV. With respect to magnetic properties, the selected free Pt$_{13}$ clusters show a magnetic character. Their total magnetization (TM), calculated as the difference between the majority and minority charge densities integrated over the real space, is reported in   \ref{Fig_selected}. The F0-F14 sample has a TM average of 5.6 $\mu_B$, with roughly the 45\% with a TM of 6 $\mu_B$. Notably, wire-like minima such as F41 and F42, which arise from the compenetration of octahedra of 6 atoms, have a very high TM (10-12 $\mu_B$). Unfortunately, these shapes are too long (R$_{max} \sim$ 8 \AA, ) to be grown into the supercage. 
The F0-F14 isomers were then inserted into the NaY zeolite at the SII supercage site, and ionically relaxed. We first note that the chemisorption is always through only one Pt vertex anchored to only one O atom of the six-ring of the supercage, in vert good agreement with experimental data. 
As depicted in   \ref{Fig_selected}, Z0 to Z14 identify the embedded clusters in zeolite using a standard DFT framework, whereas D0 to D14 refer to the case where London dispersion corrections are included. A label R indicates isomers that reconstruct significantly from their starting configuration. The relative stability of each minimum is calculated with respect to the global minimum, Z0-c and D0R-c, at DFT and DFT-D, respectively. D0R-c, which is an L-shaped motif delimited by six squares, arises from Z0-c when dispersion forces are introduced. In  \ref{Fig_zeo}, free and encaged minima are plotted as functions of their CN and TM, and the color scheme refers to their energy stability with respect to the global minimum. At the DFT level, the overall coordination of Pt isomers, and thus their magnetic properties, is almost preserved after the isomers are embedded with the exception of the unstable geometries that undergo a major reconstruction, i.e., Z6R, Z7R, Z8R, Z13R, and Z14R, as in  \ref{Fig_selected}.
Within a DFT-D framework, the net effect of adding dispersion forces is a decrease in the nearest-neighbor average distance, a preference for assuming a cage-like shape associated with a higher binding energy, and the shift towards lower coordinated isomers is made evident in  \ref{Fig_zeo}. Compared with the DFT case, the distance between the anchor and the center of the oxygen ring is always shorter by 0.1-0.3 \AA~ . Simultaneously, all the embedded NPs tend to move toward the center of the cage, where the average distance between the cluster centers of mass and the geometrical center of the pore is 0.2-0.3 \AA~ . The dispersion forces thus induce a few local atomic rearrangements that are sometimes significant. The most relevant case is the global minimum Z0-c. It evolves toward an L-shaped structure with six square facets, D0R-c, linked to the supercage through a three-fold vertex. Notably, D0R-c is unstable in the gas phase and folds again into the original LOW shape. However, its counterpart chemisorbed with a three-fold vertex, D0-b, remains geometrically stable, although it loses 0.4 eV in energy. At the DFT-D level, the $\Delta E$ is spread up to 2.5 eV and it is strongly correlated to the coordination number where $\Delta E >$ 1.5 eV are for CN > 4.7. 

\section{Discussion and Conclusion}
To fully understand the geometrical stability of zeolite-embedded Pt-NPs, we investigate the role of the vertex anchoring the cluster to the supercage. At the DFT level, the chemisorption through a 3- and 4-fold coordinated vertex usually leads to negligible changes in the overall cluster shape, whereas chemisorption through a 5-fold coordinated vertex leads to major structural transformations. 
To be more quantitative, the encapsulation energy $E_{enc}$ and the solid angle at the anchor, $\Omega$, were calculated. 
$E_{enc}$ is the energy gain due to the embedding, where values greater than 1.3 eV indicate structural rearrangement of the whole cluster.  
The values of $\Omega$ reported per each isomer in  \ref{Fig_selected} represents the angles subtended by the polygon obtained by connecting all the Pt atoms coordinated to the vertex, as projected onto the unit sphere around it. 
Our simulations show that only solid angles $\Omega$ smaller than 0.8 sr are fully stable, while values greater than 2 sr are completely unstable from a geometrical point of view; i.e., the polygon underneath the vertex/anchor can be a triangle or a quadrilateral, but not a rectangle or planar pentagon. Among clusters with $\Omega <$ 0.8 sr, the best examples are the truncated bi-pyramid (Z1) and the symmetric TTP-based structure (Z2); they have a three-fold rotational symmetry around the Pt-anchor, and a $E_{enc} \sim 0.9$ eV. A non-planar pentagon, with a borderline value of $\Omega$, is allowed, as in the case of the bilayer (Z9) which has a $E_{enc} \sim 0.9$ eV. When $\Omega >$ 2 sr and $E_{enc} >$ 1.4 eV, severe rearrangements occur to decrease $\Omega$ to approximately 1 sr, as depicted in  \ref{Fig_selected}. 
Generally speaking, structural reconstruction lowers the TM with the sole exception of the isomer arising from the Ih reconstruction. Notably, the Ih$_{13}$ indeed evolves towards a iDh$_{23}$ upon a three-fold anchor, a distorted half-star, Z4. Not surprisingly, the reconstructed Ih exhibits a TM of 4 $\mu_B$ similarly to Z4. A full description of those transformations comes as Supporting Information. Those structures that reconstructed at the DFT level because of the embedding continue to change toward less coordinated shapes at the DFT-D level. Notably, the incomplete decahedra are geometrically stable while helicoidal shapes tend to rearrange. Details of these transformations are provided at pages S5-S7 of the Supporting Information.

\begin{center}
\begin{figure}[ht!]
\includegraphics[scale=0.7]{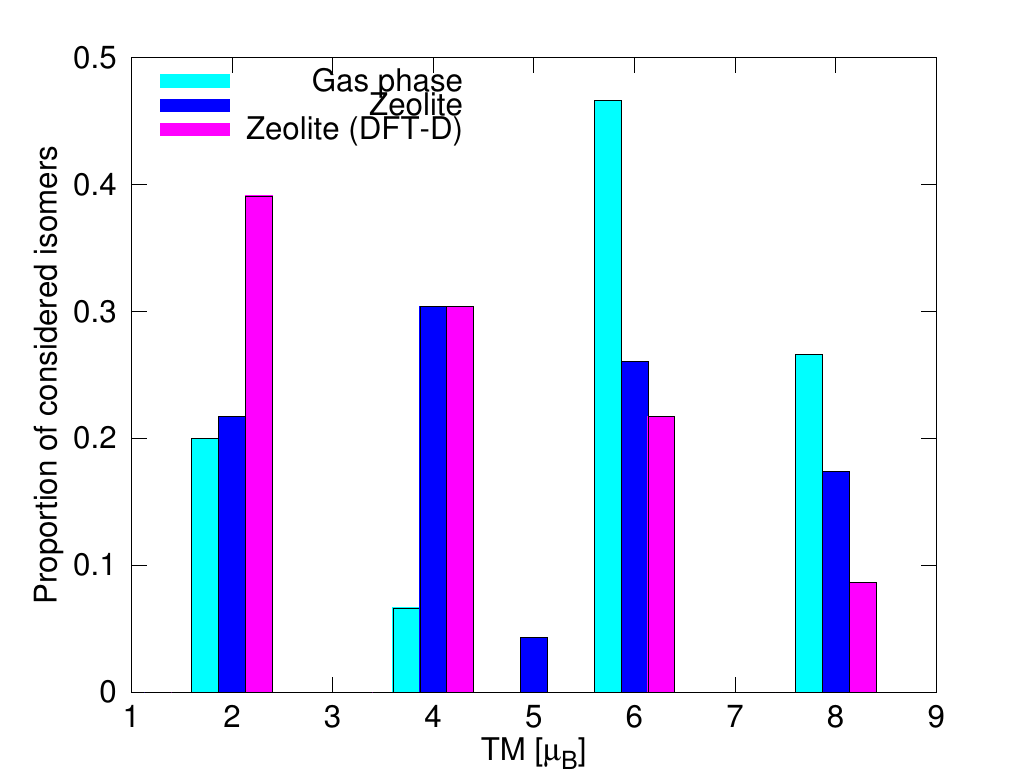}
\caption{Proportion of considered isomers in terms of their total magentization TM ($\mu_B$), in the gas phase (cyan), in zeolite (blue) and DFT-D zeolite (magenta).}
\label{Fig_magn}
\end{figure}
\end{center}

The structural rearrangements due to the embedding and the inclusion of van der Waals forces strongly affect the magnetic properties of Pt$_{13}$-NPs@NaY. Although TM ranges between 2 and 8 $\mu_B$ as in the gas phase, the average TM is about 3.9 $\mu_B$ against 4.8 $\mu_B$ at the DFT level, and 5.6 $\mu_B$ in the gas phase, as shown in  \ref{Fig_magn}. The decrease in the TM is associated to the stabilization of low coordinated shapes. D7 is the only case which increases its TM to 8 $\mu_B$ due to a different orientation within the supercage. Thus far, at the DFT-D level, only six isomers have a TM above 6 $\mu_B$~: TBP (D1);  TTP (D2); iIh$_{19}$ (D12) and the helicoidal D10-R, which have a spin state of 3; and the other two incomplete icosahedra D5 and D7 which show a high magnetization of 8 $\mu_B$. Anyway, only four of them are highly coordinated (CN>5).
Bartolom\`e and co-workers \cite{Bartolome09} suggested that the magnetic behavior of Pt-NPs could be influenced by a charge-transfer between the O atoms of the cage and the cluster. Here, we observe a charge transfer of approximately 0.2 e$^-$/atom from the O toward the Pt vertex bound to the six-ring substructure. The involved oxygen exhibits an atomic polarizability of $\pm$ 0.4 $\mu_B$ that contributes to the total magnetic moment of the whole system. The presence of two Al atoms close to the vertex/anchor of the NP results in a Pt-O distance peaked at 2.25~\AA\ in excellent agreement with experiments \cite{Liu06}.

%\section{Summary}
In conclusion, we investigated the energy landscape of Pt$_{13}$  using a metadynamics approach and found an exhaustive set of highly coordinated isomers with a maximum pair distance shorter than 7.5 \AA. Fifteen different isomers were selected and then embedded in a NaY zeolite supercage model with different orientations when possible. Each Pt$_{13}$-NP adsorbs at the SII site with a vertex anchored to the inward O atom of the six oxygen ring substructure.  The effect of the embedding was the destabilization of NPs linked to the supercage throughout solid angles larger than 2 sr, as occurs in the case of the icosahedron. Van der Waals forces were included to determine the stability of each isomer, and their magnetic properties were subsequently estimated. In the case of encapsulated Pt-NPs, a new concave structure was found to be the lowest energy structure. As in the gas phase, the embedded Pt-NPs exhibit a magnetic behavior, with a total magnetization between 2 and 8 $\mu_B$. However, due to the stabilization of low coordinated isomers, the majority exhibits a TM of 2 $\mu_B$ against a value of 6 $\mu_B$ calculated in free clusters. 
%A charge transfer between the zeolite cage and Pt atoms occured when two Al atoms occupied vicinal positions in the six-ring substructure of the zeolite, leaving the oxygen atom between them slightly polarized. 
The high magnetic moment of Pt$_{13}$-NPs in zeolite, was mainly due to four highly coordinated isomers: a truncated bi-pyramid, a triaugmented triangular prism, and two incomplete double icosahedra. Nonetheless, these isomers are quite unfavorable from an energetic point of view, which explains why only a fraction of the total Pt load is magnetically active in very good agreement with available experimental data.

{\bf Acknowledgement}.
This work is supported by the U.K. research council EPSRC, under Grant No. EP/GO03146/1. 
The authors would like to thank A. Floris and A. Comisso for their useful suggestions and R. D' Agosta for a careful reading of the manuscript. Authors acknowledge support from the COST Action MP0903 "Nanoalloys as Advanced Materials: From Structure to Properties and Applications".

{\bf Supporting Information Available}.
A full description of the adopted computational procedure is provided at pgs S1-S2. Structural features of zeolite cages and details of structural instability of embedded Pt-NPs could be found in the Supporting Information section at pgs S2-S4 with Figs S1-S2, and pgs S5-S7, respecticely. This information is available free of charge via the internet at http://pubs.acs.org/.

\bibliography{all_bib}

\newpage
\begin{center}
\begin{figure}[ht!]
\includegraphics[scale=1.0]{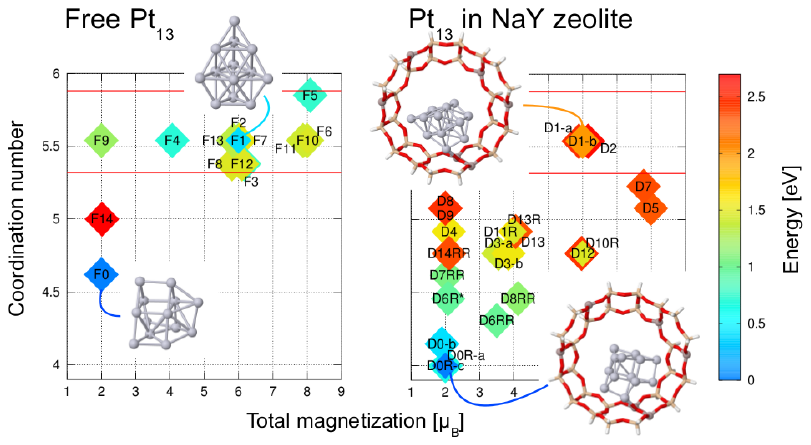}
\end{figure}
\end{center}

\end{document}